\documentstyle[12pt]{article}
\newcommand{\bfp}{\mbox{\boldmath $p$}}
\newcommand{\bfh}{\mbox{\boldmath $h$}}
\newcommand{\bfq}{\mbox{\boldmath $q$}}
\newcommand{\ZP}[1]{{\it Z.\ Phys.}\ {\bf #1}}
\newcommand{\PL}[1]{{\it Phys.\ Lett.}\ {\bf #1}}
\newcommand{\beq}{\begin{equation}}
\newcommand{\eeq}{\end{equation}}
\newcommand{\barr}{\begin{eqnarray}}
\newcommand{\earr}{\end{eqnarray}}
\newcommand{\ee}{e^-e^+}
\newcommand{\qq}{q\bar q}
\newcommand{\uu}{u\bar u}
\newcommand{\dd}{d\bar d}
\newcommand{\la}{\lambda}
\begin{document}
\begin{flushright}
DFTT 11/98 \\
hep-ph/9802435 \\
\end{flushright}
\vskip 1.5cm
\begin{center}
{\bf Spin effects in vector meson production at LEP\footnote{
Presented at Cracow Epiphany Conference on Spin Effects in Particle
Physics and TEMPUS Workshop, January 8-11, 1998}
}\\
\vskip 1.5cm
{\sf M. Anselmino}
\vskip 0.8cm
{Dipartimento di Fisica Teorica, Universit\`a di Torino and \\
INFN, Sezione di Torino, Via P. Giuria 1, 10125 Torino, Italy}
\end{center}
\vskip 1.5cm
\noindent
{\bf Abstract:}
\vspace{6pt}
\noindent
Spin observables may reveal much deeper properties of non perturbative 
hadronic physics than unpolarized quantities. We discuss the
polarization of hadrons produced in $e^+e^-$ annihilation at LEP. 
We show how final state $\qq$ interactions may give origin to non zero 
values of the off-diagonal element $\rho^{\,}_{1,-1}$ of the helicity 
density matrix of vector mesons: some predictions are given for 
$K^*, \phi, D^*$ and $B^*$ in agreement with recent OPAL data. 
We also discuss the relative amount of vector and pseudovector meson states
and the probability of helicity zero vector states. Similar measurements
in other processes are suggested.
\newpage
\pagestyle{plain}
\setcounter{page}{1}
\noindent
{\bf {\mbox{\boldmath $\rho^{\,}_{1,-1}(V)$}} in the process 
{\mbox{\boldmath $e^- e^+ \to \qq \to V + X$}}}
\vskip 12pt

The spin properties of hadrons inclusively produced in high energy 
interactions are related to the fundamental properties of quarks and 
gluons and to their elementary interactions in a much more subtle way 
than unpolarized quantities. They test unusual basic dynamical 
properties and reveal how the usual hadronization models -- 
successful in predicting unpolarized cross-sections -- may not be 
adequate to describe spin effects, say the fragmentation of a polarized 
quark.

We consider here the spin properties of hadrons created at LEP. It was pointed 
out in Refs. \cite{akp} and \cite{aamr} that final state interactions 
between the $q$ and $\bar q$ produced in $e^+ e^-$ annihilations 
-- usually neglected, but indeed necessary -- might give origin to non 
zero spin observables which would otherwise be forced to vanish: 
the off-diagonal element $\rho^{\,}_{1,-1}$ of the helicity density matrix of 
vector mesons may be sizeably different from zero \cite{akp} due to a 
coherent fragmentation process which takes into account $q \bar q$ 
interactions. The incoherent fragmentation of a single independent quark 
leads instead to zero values for such off-diagonal elements. 

We present predictions \cite{abmq} for $\rho^{\,}_{1,-1}$ of several vector 
mesons $V$ provided they are produced in two jet events, carry
a large momentum or energy fraction $z=2E_{_V}/\sqrt s$, and have a small
transverse momentum $p_{_T}$ inside the jet. Our estimates are in agreement
with the existing data and are crucially related both to the presence 
of final state interactions and to the Standard Model couplings of the 
elementary $e^- e^+ \to q \bar q$ interaction. 

The helicity density matrix of a hadron $h$ inclusively produced in the 
two jet event $e^- e^+ \to q\bar q \to h + X$ can be written as 
\cite{akp, aamr}
\beq
\rho^{\,}_{\la^{\,}_h \la^{\prime}_h}(h) 
= {1\over N_h} \sum_{q,X,\la^{\,}_X,\la^{\,}_q,\la^{\,}_{\bar q},
\la^{\prime}_q,\la^{\prime}_{\bar q}} 
D^{\,}_{\la^{\,}_h \la^{\,}_X; \la^{\,}_q,\la^{\,}_{\bar q}} \>\>
\rho^{\,}_{\la^{\,}_q,\la^{\,}_{\bar q};
\la^{\prime}_q,\la^{\prime}_{\bar q}}\,(\qq) \>\> 
D^*_{\la^{\prime}_h \la^{\,}_X; \la^{\prime}_q,\la^{\prime}_{\bar q}} 
\label{rhoh}
\eeq
where $\rho(\qq)$ is the helicity density matrix of the $q\bar q$ state 
created in the annihilation of the unpolarized $e^+$ and $e^-$,
\beq
\rho^{\,}_{\la^{\,}_q,\la^{\,}_{\bar q};
\la^{\prime}_q,\la^{\prime}_{\bar q}}\,(\qq)
= {1\over 4N_{\qq}} \sum_{\la^{\,}_{-}, \la^{\,}_{+}}
M^{\,}_{\la^{\,}_q \la^{\,}_{\bar q};\la^{\,}_{-} \la^{\,}_{+}} \>
M^*_{\la^{\prime}_q \la^{\prime}_{\bar q}; \la^{\,}_{-} \la^{\,}_{+}} \,.
\label{rhoqq}
\eeq
The $M$'s are the helicity amplitudes for the $\ee \to \qq$ process and
the $D$'s are the fragmentation amplitudes, {\it i.e.} the helicity
amplitudes for the process $\qq \to h+X$; the $\sum_{X,\lambda_X}$ stands 
for the phase space integration and the sum over spins of all the unobserved 
particles, grouped into a state $X$. The normalization factors $N_h$ and
$N_{\qq}$ are given by:
\beq
N_h = \sum_{q,X; \la^{\,}_h, \la^{\,}_X, \la^{\,}_q, \la^{\,}_{\bar q},
\la^{\prime}_q, \la^{\prime}_{\bar q}} 
D^{\,}_{\la^{\,}_h \la^{\,}_X; \la^{\,}_q,\la^{\,}_{\bar q}} \>\>
\rho^{\,}_{\la^{\,}_q,\la^{\,}_{\bar q};
\la^{\prime}_q,\la^{\prime}_{\bar q}}\,(\qq) \>\> 
D^*_{\la^{\,}_h \la^{\,}_X; \la^{\prime}_q,\la^{\prime}_{\bar q}} \,
= \sum_q D^h_q \,,
\label{nh}
\eeq
where $D^h_q$ is the usual fragmentation function of quark $q$ into 
hadron $h$, and 
\beq
N_{\qq} = {1\over 4} 
\sum_{\la^{\,}_q, \la^{\,}_{\bar q}; \la^{\,}_{-}, \la^{\,}_{+}} \vert 
M^{\,}_{\la^{\,}_q \la^{\,}_{\bar q}; \la^{\,}_{-} \la^{\,}_{+}} \vert^2 \,.
\label{nqq}
\eeq

The helicity density matrix for the $q\bar q$ state can be computed 
in the Standard Model and its non zero elements are given by
\barr
\rho^{\,}_{+-;+-}(\qq) &=& 1 - \rho^{\,}_{-+;-+}(\qq) \> \simeq \> 
{1\over 2}\,{(g_{_V} - g_{_A})^2_q \over (g^2_{_V} + g^2_{_A})_q}
\label{rhoqqd} \\
\rho^{\,}_{+-;-+}(\qq) &=& \rho^*_{+-;-+}(\qq) \> \simeq \> 
{1\over 2}\,{(g^2_{_V} - g^2_{_A})_q \over (g^2_{_V} + g^2_{_A})_q} 
\, {\sin^2\theta \over 1+ \cos^2\theta} \, \cdot
\label{rhoqqap}
\earr
These expressions are simple but approximate and hold at the $Z_0$ pole, 
neglecting electromagnetic contributions, masses and terms proportional 
to $g_{_V}^l$; the full correct expressions can be found in Ref. \cite{abmq}.

Notice that, inserting the values of the coupling constants
\barr
g_{_V}^{u,c,t} &=&  \>\> {1\over 2} - {4\over 3}\sin^2\theta_{_W} \quad\quad
g_{_A}^{u,c,t} = {1\over 2} \label{cc} \\
g_{_V}^{d,s,b} &=& -{1\over 2} + {2\over 3}\sin^2\theta_{_W} \quad\quad
g_{_A}^{d,s,b} = -{1\over 2} 
\nonumber
\earr 
one has
\barr
\rho^{\,}_{+-;-+}(\uu, c\bar c, t\bar t) 
&\simeq& -0.36 {\sin^2\theta \over 1 + \cos^2\theta} \label{rho+-ap} \\
\rho^{\,}_{+-;-+}(\dd, s\bar s, b\bar b) 
&\simeq& -0.17 {\sin^2\theta \over 1 + \cos^2\theta} 
\, \cdot \nonumber
\earr

Eq. (\ref{rho+-ap}) clearly shows the $\theta$ dependence of 
$\rho^{\,}_{+-;-+}(\qq)$. In case of pure electromagnetic interactions
($\sqrt s \ll M_{_Z}$) one has exactly:
\beq
\rho^{\gamma}_{+-;-+}(\qq) = {1\over 2}
\,{\sin^2\theta \over 1+ \cos^2\theta} \, \cdot
\label{rhoqqelm}
\eeq
Notice that Eqs. (\ref{rho+-ap}) and (\ref{rhoqqelm}) have the same
angular dependence, but a different sign for the coefficient in front,
which is negative for the $Z$ contribution.

By using the above equations for $\rho(\qq)$ into Eq. (\ref{rhoh}) one 
obtains the most general expression of $\rho(h)$ in terms of the $\qq$ spin 
state and the unknown fragmentation amplitudes.

Despite the ignorance of the fragmentation process some predictions
can be made \cite{abmq} by considering the production of hadrons almost 
collinear with the parent jet: the $q \bar q \to h + X$ fragmentation is 
then essentially a c.m. forward process and the unknown $D$ amplitudes must 
satisfy the angular momentum conservation relation \cite{bls}
\beq
D^{\,}_{\la^{\,}_h \la^{\,}_X; \la^{\,}_q,\la^{\,}_{\bar q}} 
\sim \left( \sin{\theta_h\over 2} \right)^{
\vert \la^{\,}_h - \la^{\,}_X - \la^{\,}_q + \la^{\,}_{\bar q} \vert} \,,
\label{frd}
\eeq
where $\theta_h$ is the angle between the hadron momentum, 
$\bfh = z \bfq + \bfp_{_T}$, and the quark momentum $\bfq$, that is
\beq \sin\theta_h \simeq {2p_{_T} \over z\sqrt s} \,\cdot \label{ptb} \eeq
The bilinear combinations of fragmentation amplitudes contributing to
$\rho(h)$ are then not suppressed by powers of $(p_{_T}/z\sqrt s)$
only if the exponent in Eq. (\ref{frd}) is zero, which greatly reduces the 
number of relevant helicity configurations.

The fragmentation process is a parity conserving one and the fragmentation
amplitudes must then also satisfy the forward parity relationship
\beq
D^{\,}_{-\la^{\,}_h -\la^{\,}_X; -+} = (-1)^{S^{\,}_h + S^{\,}_X + 
\la^{\,}_h - \la^{\,}_X} \> 
D^{\,}_{\la^{\,}_h \la^{\,}_X; +-} \,.
\label{par}
\eeq

Before presenting analytical and numerical results for the coherent quark 
fragmentation let us remember that in case of incoherent single quark 
fragmentation Eq. (\ref{rhoh}) becomes
\beq
\rho^{\,}_{\la^{\,}_h \la^{\prime}_h}(h) 
= {1\over N_h} \sum_{q,X,\la^{\,}_X,\la^{\,}_q,\la^{\prime}_q}
D^{\,}_{\la^{\,}_h \la^{\,}_X; \la^{\,}_q} \>\>
\rho^{\,}_{\la^{\,}_q \la^{\prime}_q} \>\>
D^*_{\la^{\,}_h \la^{\,}_X; \la^{\,}_q} \,,
\label{rhohp1}
\eeq
where $\rho(q)$ is the quark $q$ helicity density matrix related to
$\rho(\qq)$ by
\beq
\rho^{\,}_{\la^{\,}_q \la^{\prime}_q} = \sum_{\la^{\,}_{\bar q}}
\rho^{\,}_{\la^{\,}_q, \la^{\,}_{\bar q}; \la^{\prime}_q,
\la^{\,}_{\bar q}} (\qq) \,.
\eeq

In such a case angular momentum conservation for the collinear quark 
fragmentation requires $\la^{\,}_q = \la^{\,}_h + \la^{\,}_X$;
the Standard Model computation of $\rho(q)$ gives only non zero diagonal
terms [$\rho^{\,}_{++}(q) = \rho^{\,}_{+-;+-}(\qq)$ and 
$\rho^{\,}_{--}(q) = \rho^{\,}_{-+;-+}(\qq)$], and one ends up
with the usual probabilistic expression
\beq
\rho^{\,}_{\la^{\,}_h \la^{\,}_h}(h) 
= {1\over N_h} \sum_{q, \la^{\,}_q}
\rho^{\,}_{\la^{\,}_q \la^{\,}_q} \>\>
D_{q,\la^{\,}_q}^{h, \la^{\,}_h} \,,
\label{rhohp2}
\eeq
where $D_{q,\la^{\,}_q}^{h, \la^{\,}_h}$ is the polarized fragmentation 
function of a $q$ with helicity $\la^{\,}_q$ into a hadron $h$ with 
helicity $\la^{\,}_h$. Off-diagonal elements of $\rho(h)$ are all zero.

\vskip 12 pt
\noindent
{\bf 2. {\mbox{\boldmath $e^- e^+ \to BX \> (S_{_B} = 1/2, 
\> p_{_T}/\sqrt s \to 0)$}}}
\vskip 12pt

Let us consider first the case in which $h$ is a spin 1/2 baryon. It was
shown in Ref. \cite{aamr} that in such a case the coherent quark 
fragmentation only induces small corrections to the usual incoherent 
description 
\barr
\rho^{\,}_{++}(B) &=& {1 \over N_{_B}} \sum_q 
\left[ \rho^{\,}_{+-;+-}(\qq) \> D_{q,+}^{B,+} + \rho^{\,}_{-+;-+}(\qq) \> 
D_{q,-}^{B,+} \right] \\
\rho^{\,}_{+-}(B) &=& {\cal O} \left[ \left(
{p_{_T} \over z \sqrt s} \right) \right] \label{rho+-b}\,.
\earr

That is, the diagonal elements of $\rho(B)$ are the same as those given by 
the usual probabilistic formula (\ref{rhohp2}), with small corrections
of the order of $(p_{_T}/z\sqrt s)^2$, while off-diagonal elements are 
of the order $(p_{_T}/z\sqrt s)$ and vanish in the $p_{_T}/\sqrt s \to 0$
limit.

The matrix elements of $\rho(B)$ are related to the longitudinal ($P_z$)
and transverse ($P_y$) polarization of the baryon:
\beq
P_z = 2\rho_{++} - 1, \quad\quad\quad\quad P_y = -2\,{\rm Im} \rho_{+-} \,.
\eeq
Some data are available on $\Lambda$ polarization, both longitudinal and
transverse, from ALEPH Collaboration \cite{aleph} and they do agree with the
above equations. In particular the transverse polarization, at 
$\sqrt s = M_{_Z}$, $p_{_T} \simeq 0.5$ GeV/$c$ and $z \simeq 0.5$ is 
indeed of the order 1\%, as expected from Eq. (\ref{rho+-b}).

\vskip 12pt
\noindent
{\bf 3. {\mbox{\boldmath $e^- e^+ \to VX \> (S_{_V} = 1, 
\> p_{_T}/\sqrt s \to 0)$}}}
\vskip 12pt

In case of final spin 1 vector mesons one has, always in the limit of small
$p_{_T}$ \cite{akp}, \cite{abmq} 
\barr
\rho^{\,}_{00}(V) &=& {1 \over N_{_V}} \sum_q D_{q,+}^{V,0} \label{rhod} \\
\rho^{\,}_{11}(V) &=& {1 \over N_{_V}} \sum_q
\left[ \rho_{+-;+-}(\qq) D_{q,+}^{V,1} + \rho_{-+;-+}(\qq) D_{q,-}^{V,1}
\right] \\
\rho^{\,}_{1,-1}(V) &=& {1 \over N_{_V}} \sum_{q,X}
D^{\,}_{10;+-} \> D^*_{-10;-+} \> \rho^{\,}_{+-;+-}(\qq) \,.
\earr

Again, the diagonal elements have the usual probabilistic expression; 
however, there is now an off-diagonal element, $\rho^{\,}_{1,-1}$, 
which may survive even in the $p_{_T}/\sqrt s \to 0$ limit. In the sequel
we shall concentrate on it. Let us first notice that, in the collinear limit,
one has
\barr
D^{V,0}_{q,+} &=& \sum_X \vert D^{\,}_{0-1;+-} \vert^2 = D^{V,0}_{q,-} \\
D^{V,1}_{q,+} &=& \sum_X \vert D^{\,}_{10;+-} \vert^2 = D^{V,-1}_{q,-} \\
D^{V,1}_{q,-} &=& \sum_X \vert D^{\,}_{12;-+} \vert^2 = D^{V,-1}_{q,+} \,,
\earr
with $ D_q^V = D^{V,0}_{q,+} + D^{V,1}_{q,+} + D^{V,-1}_{q,+}$ and 
$N_{_V} = \sum_q D_q^V$. We also notice that the two fragmentation 
amplitudes appearing in Eq. (21) are related by parity and their product
is always real. $\rho^{\,}_{00}$ and $\rho^{\,}_{1,-1}$ can be measured
through the angular distribution of two body decays of $V$. 

In order to give numerical estimates of $\rho^{\,}_{1,-1}$ we make some
plausible assumptions
\barr
D^{h,1}_{q,-} &=& D^{h,-1}_{q,+} = 0 \label{ass1} \\
D^{h,0}_{q,+} &=& \alpha^V_q \> D^{h,1}_{q,+} \label{ass2} \,.
\earr
The first of these assumptions simply means that quarks with helicity
1/2 ($-1/2$) cannot fragment into vector mesons with helicity $-1$ ($+1$).
This is true for valence quarks assuming vector meson wave functions 
with no orbital angular momentum, like in $SU(6)$. The second assumption 
is also true in $SU(6)$ with $\alpha^V_q = 1/2$ for 
any valence $q$ and $V$. Rather than taking  
$\alpha^V_q = 1/2$ we prefer to relate the value of $\alpha^V_q$ to the 
value of $\rho^{\,}_{00}(V)$ which can be or has been measured.
In fact, always in the $p_{_T} \to 0$ limit, one has \cite{abmq}
\beq
\rho^{\,}_{00}(V) = {\sum_q \alpha^V_q \, D^{h,1}_{q,+}
\over \sum_q \> (1+\alpha^V_q) \, D^{h,1}_{q,+}} \,\cdot
\label{rho00}
\eeq
If $\alpha^V_q$ is the same for all valence quarks in $V$ 
($\alpha^V_q = \alpha^V$) 
one has, for the valence quark contribution:
\beq
\alpha^V = {\rho^{\,}_{00}(V) \over 1 - \rho^{\,}_{00}(V)} \,\cdot
\label{alrho}
\eeq

Finally, one obtains \cite{abmq}
\beq
\rho^{\,}_{1,-1}(V) \simeq [1 - \rho^{\,}_{0,0}(V)] \,
{\sum_q \, D^{V,1}_{q,+} \> \rho_{+-;-+}(\qq) 
\over \sum_q \, D^{V,1}_{q,+}} \,\cdot
\label{rho1-1tss}
\eeq

We shall now consider some specific cases in which we expect 
Eq. (\ref{rho1-1tss}) to hold; let us remind once more that our 
conclusions apply to spin 1 vector mesons produced in 
$e^- e^+ \to q \bar q \to V+X$ processes in the limit of small $p_{_T}$
and large $z$, {\it i.e.}, to vector mesons produced in two jet events
($e^- e^+ \to \qq$) and collinear with one of them ($p_{_T} = 0$), 
which is the jet generated by a quark which is a valence quark for the
observed vector meson (large $z$). These conditions should be met 
more easily in the production of heavy vector mesons. 

One obtains \cite{abmq}:
\barr
\rho^{\,}_{1,-1}(B^{*}) &\simeq& [1 - \rho^{\,}_{0,0}(B^{*})] \>
\rho_{+-;-+}(b \bar b) \label{rhoba} \\
\rho^{\,}_{1,-1}(D^{*}) &\simeq& [1 - \rho^{\,}_{0,0}(D^{*})] \>
\rho_{+-;-+}(c \bar c) \label{rhoda} \\
\rho^{\,}_{1,-1}(\phi)  &\simeq& [1 - \rho^{\,}_{0,0}(\phi)] \>
\rho_{+-;-+}(s\bar s) \label{rhopa} \\
\rho^{\,}_{1,-1}(\rho) &\simeq& {1\over 2} \> [1 - \rho^{\,}_{0,0}(\rho)] \>
[\rho_{+-;-+}(u\bar u) + \rho_{+-;-+}(d \bar d)] \label{rhora} \\
\rho^{\,}_{1,-1}(K^{*\pm}) &\simeq& {1\over 2} \> 
[1 - \rho^{\,}_{0,0}(K^{*\pm})] 
\> [\rho_{+-;-+}(u\bar u) + \rho_{+-;-+}(s\bar s)] \label{rhpk+a} \\
\rho^{\,}_{1,-1}(K^{*0}) &\simeq& {1\over 2} \> [1 - \rho^{\,}_{0,0}(K^{*0})] 
\> [\rho_{+-;-+}(d\bar d) + \rho_{+-;-+}(s\bar s)] \,. \label{rhok0a} 
\earr

Eqs. (\ref{rhoba})-(\ref{rhok0a}) show how the value of $\rho^{\,}_{1,-1}(V)$
are simply related to the off-diagonal helicity density matrix element 
$\rho_{+-;-+}(\qq)$ of the $\qq$ pair created in the elementary 
$e^- e^+ \to \qq$ process; such off-diagonal elements would not appear 
in the incoherent independent fragmentation of a single quark, yielding 
$\rho^{\,}_{1,-1}(V)=0$.

By inserting into the above equations the value of $\rho^{\,}_{00}$ when
available \cite{opal} and the expressions of $\rho^{\,}_{+-;-+}$,
Eq. (8), one has:
\barr
\rho^{\,}_{1,-1}(B^{*}) &\simeq& -(0.109 \pm 0.015) \ 
{\sin^2\theta \over 1 + \cos^2\theta} \\
\rho^{\,}_{1,-1}(D^{*}) &\simeq& -(0.216 \pm 0.007) \ 
{\sin^2\theta \over 1 + \cos^2\theta} \\
\rho^{\,}_{1,-1}(\phi) &\simeq& -(0.078 \pm 0.014) \ 
{\sin^2\theta \over 1 + \cos^2\theta} \\
\rho^{\,}_{1,-1}(K^{*0}) &\simeq& -0.170 \ 
[1- \rho^{\,}_{0,0}(K^{*0})] \ {\sin^2\theta \over 1 + \cos^2\theta}  \,\cdot
\earr
Finally, in case one collects all meson produced at different angles in
the full available $\theta$ range (say $\alpha < \theta < \pi -\alpha, 
\> |\cos\theta| < \cos\alpha$) an average should be taken in $\theta$, 
weighting the different values of $\rho^{\,}_{1,-1}(\theta)$ with the 
cross-section for the $e^-e^+ \to V+X$ process; this gives \cite{abmq}:
\barr
\langle \rho^{\,}_{1,-1}(B^{*}) \rangle_{[\alpha, \pi-\alpha]}
&\simeq& -(0.109 \pm 0.015) \ 
{3 - \cos^2\alpha \over 3 + \cos^2\alpha} \label{rhobn} \\
\langle \rho^{\,}_{1,-1}(D^{*}) \rangle_{[\alpha, \pi-\alpha]}
&\simeq& -(0.216 \pm 0.007) \ 
{3 - \cos^2\alpha \over 3 + \cos^2\alpha} \label{rhodn} \\
\langle \rho^{\,}_{1,-1}(\phi) \rangle_{[\alpha, \pi-\alpha]}
&\simeq& -(0.078 \pm 0.014) \ 
{3 - \cos^2\alpha \over 3 + \cos^2\alpha} \label{rhopn} \\
\langle \rho^{\,}_{1,-1}(K^{*0}) \rangle_{[\alpha, \pi-\alpha]}
&\simeq& -0.170 \ [1- \rho^{\,}_{0,0}(K^*)] \ 
{3 - \cos^2\alpha \over 3 + \cos^2\alpha} \,\cdot \label{rhok0n} 
\earr

These results have to be compared with data \cite{opal}
\barr
\rho^{\,}_{1,-1}(D^*) &=& -0.039 \pm 0.016 \quad\quad {\rm for}
\quad\quad z > 0.5 \quad\quad \cos\alpha = 0.9 \\ 
\rho^{\,}_{1,-1}(\phi) &=& -0.110 \pm 0.070 \quad\quad {\rm for}
\quad\quad z > 0.7 \quad\quad \cos\alpha = 0.9 \\ 
\rho^{\,}_{1,-1}(K^{*0}) &=& -0.090 \pm 0.030 \quad\quad {\rm for}
\quad\quad z > 0.3 \quad\quad \cos\alpha = 0.9 
\earr
which shows a good qualitative agreement with the theoretical
predictions. We notice that while the mere fact that $\rho_{1,-1}$ differs 
from zero is due to a coherent fragmentation of the $\qq$ pair, the actual 
numerical values depend on the Standard Model coupling constants; for example,
$\rho_{1,-1}$ would be positive at smaller energies, at which the one gamma
exchange dominates, while it is negative at LEP energy where the one $Z$
exchange dominates. $\rho_{1,-1}$ has also a peculiar dependence on 
the meson production angle, being small at small and large angles
and maximum at $\theta = \pi/2$. Such angular dependence has been tested
in case of $K^{*0}$ production and indeed one has \cite{opal}, in agreement 
with Eqs. (\ref{rhok0a}) and (\ref{rho+-ap}),
\beq
\left[ {\rho^{\,}_{1,-1} \over 1- \rho^{\,}_{00}} \right]_{|\cos\theta|<0.5} 
\cdot 
\left[ {\rho^{\,}_{1,-1} \over 1- \rho^{\,}_{00}} \right]^{-1}
_{|\cos\theta|>0.5} = 1.5 \pm 0.7 \,\,.
\eeq

\vskip 12pt
\noindent
{\bf 4. Diagonal elements of {\mbox{\boldmath $\rho(V)$}} and 
{\mbox{\boldmath $P_{_V} \equiv V/(V+P)$}}}
\vskip 12pt

Let us consider now the diagonal element $\rho_{00}(V)$ -- for which
the probabilistic interpretation, Eq. (\ref{rhod}) holds -- together with 
the production of pseudoscalar mesons; that is, we consider the production
of the pseudoscalar mesons $P = K, D, B$ and the corresponding vector mesons 
$V = K^*, D^*, B^*$: data are available on $\rho_{00}(V)$ and the ratio of 
vector to vector + pseudoscalar mesons, $P_{_V} \equiv V/(V+P)$ 
\cite{opal, opal2}. 

We denote by $P^{\la}_S$ the probability that the fragmenting quark produces 
a meson with spin $S$ and helicity $\la$ and consider only the production of 
vector and pseudovector mesons. Then we have \cite{burg, rhovp}:
\beq
\rho_{00}(V) = {P_1^0 \over P_1^{\pm1} + P_1^0} \quad\quad\quad
P_V = P_1^{\pm1} + P_1^0 
\eeq
with
$P_1^{\pm1} = P_1^1 + P_1^{-1}$ and $P_1^{\pm1} + P_1^0 + P_0^0 = 1$.

In terms of the fragmentation functions this reads:
\beq
P_0^0 = {D_q^P \over D_q^V + D_q^P} \quad\quad
P_1^0 = {D_q^{V,0} \over D_q^V + D_q^P} \quad\quad
P_1^{\pm1} = {D_q^{V,1} + D_q^{V,-1} \over D_q^V + D_q^P} 
\eeq
Notice that, by parity invariance, the above quantities are independent
of the quark helicity.  

Statistical spin counting would give
\beq
P_1^{\pm1} = 0.5 \quad\quad P_1^0 = 0.25 \quad\quad P_0^0 = 0.25
\label{stat}
\eeq
that is
\beq
\rho_{00} = {1 \over 3} \quad\quad P_V = {3 \over 4}
\eeq
so that the vector meson alignment is zero:
\beq
A = {1 \over 2} \> (3 \rho_{00} - 1) = 0
\label{ali}
\eeq

From data \cite{opal2} one obtains \cite{rhovp} 
\beq
P_1^{\pm1}(K^*) = 0.34 \pm 0.06 \quad\quad
P_1^0(K^*) = 0.41 \pm 0.07 \quad\quad
P_0^0(K) = 0.25 \pm 0.10 \label{res1}
\eeq
\beq
P_1^{\pm1}(D^*) = 0.34 \pm 0.04 \quad\quad
P_1^0(D^*) = 0.23 \pm 0.03 \quad\quad
P_0^0(D) = 0.43 \pm 0.06 \label{res2}
\eeq
\beq
P_1^{\pm1}(B^*) = 0.49 \pm 0.09 \quad\quad
P_1^0(B^*) = 0.27 \pm 0.08 \quad\quad
P_0^0(B) = 0.24 \pm 0.09 \label{res3}
\eeq

The simultaneous measurements of the ratio of vector to vector + pseudovector 
mesons and $\rho_{00}(V)$ supply basic information on the fragmentation of 
quarks which does not depend on the helicity of the quark, but 
on the spin and helicity of the final meson. The data available for $K$, $D$ 
and $B$ mesons show clear deviations from simple statistical spin counting; 
such information could be of crucial importance for the correct formulation
of quark fragmentation Monte Carlo programs, which at the moment
widely assume simple relative statistical probabilities.

Let us consider our results, Eqs. (\ref{res1})-(\ref{res3}).
For strange mesons the data agree with spin counting in the amount of
$K$ versus $K^*$, but, among vector mesons, helicity zero states seem
to be favoured; these are in absolute the most abundantly produced,
$P_1^0(K^*) = 0.41$. For charmed mesons results differ from the spin 
counting values (\ref{stat}), suggesting a prevalence of pseudoscalar states, 
$P_0^0(D) = 0.43$, and, among vector mesons, of helicity 0 states.
The heavy $b$-mesons, instead, are produced in good agreement with
statistical spin counting rules, as one expects.

\vskip 12pt
\noindent
{\bf 5. {\mbox{\boldmath $\rho_{1,-1}(V)$}} in other processes
and conclusions} 
\vskip 12pt

The results discussed here are encouraging; indeed measurements 
of off-diagonal and diagonal elements of $\rho(V)$ give valuable
information on the hadronization process and test the underlying 
elementary dynamics. It would be very helpful to have more and 
more detailed data, possibly with a selection of final hadrons with the 
required features for our results to hold. 

It would be interesting to test the coherent fragmentation of quarks in 
other processes \cite{pir}, like $\gamma\gamma \to VX$, $pp \to D^*X$ and 
$\gamma p \to VX$ or $\gamma^* p \to VX$. The first two processes are 
similar to $e^-e^+ \to VX$ in that a $\qq$ pair is created which then 
fragments coherently into the observed vector meson; one assumes
that the dominating elementary process in $pp \to D^*X$ is $gg \to c \bar c$.
In both these cases one has for $\rho^{\,}_{+-;-+}(\qq)$ the same value
as in Eq. (\ref{rhoqqelm}), so that one expects a {\it positive} value
of $\rho^{\,}_{1,-1}(V)$. 

In the case of the real photo-production of vector mesons the quark 
fragmentation is in general a more complicated interaction of the struck 
quark with the remnants of the proton and it might be more difficult to 
obtain numerical predictions. However, if one observes $D^*$ mesons one 
can assume or select kinematical regions for which the underlying elementary 
interaction is $\gamma g \to c\bar c$: again, one would have the same 
$\rho^{\,}_{+-;-+}(c\bar c)$ as in Eq. (\ref{rhoqqelm}), and one would 
expect a positive value of $\rho^{\,}_{1,-1}(D^*)$. Similarly for the 
production of $\phi$ or $B^*$. 

The production of vector mesons like $D^*$ or $B^*$ in DIS is even more 
interesting; the polarization of the virtual photon depends on $x$ and 
$Q^2$ and so does the value of $\rho^{\,}_{+-;-+}(c\bar c)$ \cite{pir}. 
An eventual dependence of $\rho^{\,}_{1,-1}(D^*)$ on $x$ and $Q^2$
would then be an unambigous test of the hadronization mechanism and the 
elementary interaction. A similar situation can be obtained by considering
$e^+e^-$ or $\gamma\gamma$ processes with polarized initial particles:
in such cases the value of $\rho^{\,}_{+-;-+}(\qq)$ strongly depends
on the initial spin states and should change the measured value of
$\rho^{\,}_{1,-1}(V)$ \cite{new}.

Similar considerations hold for the diagonal elements of $\rho(V)$ 
and for $P_{_V}$; the degree of universality of quark fragmentation could 
be tested by studying these same quantities in other processes, like the
ones mentioned above. It would also be interesting to compare data on the 
production of spin 1/2 and spin 3/2 baryons. 

To conclude, some non perturbative aspects of strong interactions can only 
be tackled by gathering experimental information and looking for patterns and 
regularities which might allow the formulation of correct phenomenological 
models. Spin dependent quantities are still in a first stage of consideration
and development, so that even qualitative studies are meaningful; for 
example, it would indeed be interesting to perform the simple tests of 
coherent fragmentation effects suggested here.

\vskip 24pt
\noindent
{\bf Acknowledgements}
\vskip 6pt
I would like to thank the organizers of the Conference for their work and 
most successful efforts. I acknowledge financial support from
the TEMPUS program MJEP 9006-95.

\end{document}